\documentstyle[emulateapj]{article}

\begin{document}

\title{Northern Sky Variability Survey (NSVS): Public data release.\footnotemark[1]}

\author{
P. R. Wo\'zniak\altaffilmark{2},
W. T. Vestrand\altaffilmark{2},
C. W. Akerlof\altaffilmark{3},
R. Balsano\altaffilmark{2},
J. Bloch\altaffilmark{2},
D. Casperson\altaffilmark{2},
S. Fletcher\altaffilmark{2},
G. Gisler\altaffilmark{2},
R. Kehoe\altaffilmark{3, 4},
K. Kinemuchi\altaffilmark{4},
B. C. Lee\altaffilmark{5},
S. Marshall\altaffilmark{6},
K. E. McGowan\altaffilmark{2},
T. A. McKay\altaffilmark{3},
E. S. Rykoff\altaffilmark{3},
D. A. Smith\altaffilmark{3},
J. Szymanski\altaffilmark{2},
J. Wren\altaffilmark{2}
}

\vspace{0.5cm}

\footnotetext[1]{Based on observations obtained with the ROTSE-I
Robotic Telescope which was operated at Los Alamos National Laboratory.}

\altaffiltext{2}{Los Alamos National Laboratory, MS-D436, Los Alamos, NM 87545; wozniak@lanl.gov}
\altaffiltext{3}{Department of Physics, 2477 Randall Laboratory, University of Michigan,
                                  Ann Arbor, MI 48109}
\altaffiltext{4}{Department of Physics and Astronomy, Michigan State University,
                        East Lansing, MI 48824-2320}
\altaffiltext{5}{Lawrence Berkeley National Laboratory, Berkeley, CA 94720-8160}
\altaffiltext{6}{Lawrence Livermore National Laboratory, Livermore, CA 94550}

\begin{abstract}

The Northern Sky Variability Survey (NSVS) is a temporal record of the sky over the optical magnitude
range from 8 to 15.5. It was conducted in the course of the first generation Robotic Optical Transient
Search Experiment (ROTSE-I) using a robotic system of four co-mounted unfiltered telephoto lenses equipped
with CCD cameras. The survey was conducted from Los Alamos, NM, and primarily covers the entire northern sky.
Some data in southern fields between declinations $0\arcdeg$ and $-38\arcdeg$ is also available, although
with fewer epochs and noticeably lesser quality. The NSVS contains light curves for approximately 14 million
objects. With a one year baseline and typically 100--500 measurements per object, the NSVS is the most
extensive record of stellar variability across the bright sky available today. In a median field, bright
unsaturated stars attain a point to point photometric scatter of $\sim$0.02 mag and position errors
within 2$\arcsec$. At Galactic latitudes $|b|<20\arcdeg$ the data quality is limited by severe blending
due to $\sim$14$\arcsec$ pixel size. We present basic characteristics of the data set and describe data collection,
analysis, and distribution. All NSVS photometric measurements are available for on-line public access from
the Sky Database for Objects in Time-Domain (SkyDOT; {\it http://skydot.lanl.gov}) at LANL. Copies of the full
survey photometry may also be requested on tape.

\end{abstract}


\section{Introduction}

An amazing fact about modern astronomy is that the global time variability of the optical sky is largely
unexplored for objects fainter than those observable with the naked eye (Paczy\'nski 1997). As a result,
the existing samples of known variables are quite incomplete.  The  commonly accepted standard catalog
of variable stars is the General Catalog of Variable Stars (GCVS; Kholopov 1998).
However, the GCVS catalog  was compiled from a multitude of heterogeneous
observational sources---in many cases based on analysis of photographic plates---
and does not present any light curves. Spatial coverage in the GCVS is
strikingly patchy at magnitude 10 and fainter. With the data currently
available for most stars, we cannot answer the simple question of
whether or not the  star is variable. This is
unfortunate because temporal flux changes not only carry useful
physical information about the star but they are of concern for experiments where variability could degrade the quality
of the comparison star grid (e.g. Space Interferometry Mission, Frink et al. 2001).

The reasonably sensitive, inexpensive CCDs that have become available in the last few years,
coupled with the improved affordable data processing capabilities, have opened new windows
for discovery in astrophysics. The large, uniform samples being collected with CCD imagers are enabling the temporal
study of objects in new detail. Microlensing surveys, for example, have shown that massive photometric monitoring
programs can return numerous scientific results, often unrelated to the original goal (Paczy\'nski 2000a,
Ferlet, Maillard \& Raban, eds. 1997). The added value of large number statistics and good sky coverage
is also evident in catalogs from digitized POSS plates (Djorgovski et al. 2001) and multicolor surveys like
SDSS (Stoughton et al. 2002, Abazjian et al. 2003) and 2MASS (Skrutskie et al. 1997) when applied to rare
objects such as high-${\it z}$ quasars and galaxies or brown dwarfs. 

A new generation of small robotic sky patrol instruments is making all sky temporal monitoring of point sources possible
(Paczy\'nski 2000b, Chen, Lemme \& Paczy\'nski, eds. 2001). New results from these variability studies are contributing
to our understanding of stellar evolution and Galactic structure. Resulting improvements in the local distance scale
and discoveries of supernovae enable more accurate estimates of cosmological parameters and stellar ages.
A substantial contribution to the understanding of gamma-ray bursts (GRBs) has already been made by robotic
follow-up telescopes of modest size like ROTSE (Akerlof et al. 2000b, Kehoe et al. 2001)
and LOTIS (Park et al. 2002). Development of autonomous systems searching for optical flashes in real time will enable
monitoring of a variety of fast and rare phenomena including the onset of optical emission from GRBs. The RAPTOR system
(Vestrand et al. 2002) is a stereoscopic sky monitoring system that is designed to find such flashes and provide
instant notification and response while events are occurring. In the context of extra-solar planets, a shallow
but very large area time domain survey with high cadence and sub-percent photometry currently offers the best
prospect of discovering bright systems with transiting Jovian planets that are suitable for detailed studies using
high resolution spectroscopy (Horne 2002, Charbonneau 2002).

To date, the largest volumes of data on time variability are those collected by microlensing searches,
containing typically $10^7$ light curves, each with a few hundred individual photometric measurements spread
over several years (Paczy\'nski 2001, Brunner et al. 2001). Microlensing surveys can be considered medium
angle surveys limited to specific areas of interest, namely the Galactic Bulge and several galaxies of the
Local Group. In the category of very wide field surveys, the All Sky Automated Survey (ASAS, Pojma\'nski 1997)
has made a substantial fraction of the data available in the public domain and returned $\sim$7,000 variable stars
brighter than 15 mag discovered primarily in the 0--6$^h$ quadrant and additional scattered fields of the southern
hemisphere (Pojma\'nski 2000, 2002). The search for transits by extra-solar Jupiters resulted in recent proliferation
of large area photometric monitoring projects (Horne 2003), however little data has been published to date.

This paper marks the first release of data from the Northern Sky Variability Survey (NSVS), a CCD-based
synoptic survey covering the entire sky north of the declination $\delta=-38\arcdeg$. This data
release provides light curves for $\sim$14 million objects down to $V\sim$15.5 magnitude, with hundreds of repeated
observations spanning one full year. NSVS photometry is a great improvement over photographic work despite
disadvantages of using a single unfiltered photometric band. Most importantly, all observations were collected with
the same instrument resulting in uniform data quality limited
largely by crowding near the Galactic plane. We are working toward making the NSVS the best possible tool for studies
of stellar variability and Galactic structure using bright stars. Akerlof et al. (2000a) published a preliminary
variability analysis covering 9 out of 161 survey tiles and 3 months of observing time. While we repeat parts of their
discussion of the observing system and data reduction, there are significant differences in data processing between this
work and Akerlof et al. (2000a).

The outline of the paper is as follows. Section~\ref{sec:instruments} describes the telescope, cameras and the
process of data collection. Section~\ref{sec:processing} presents the details of data reductions and photometry
followed by the discussion of survey quality and coverage. In Section~\ref{sec:data} we describe public access
to the data and some technicalities of the data products. Section~\ref{sec:discussion} concludes with the summary
and future prospects for NSVS.

\section{Instruments and Data Acquisition}
\label{sec:instruments}

\noindent
Numerical data for this section is summarized in Table~\ref{tab:specs}.

\subsection{ROTSE-I Robotic Telescope}
\label{sec:telescope}

All data in the present NSVS dataset were collected by the first generation Robotic Optical Transient Search Experiment (ROTSE-I).
The primary goal of that experiment was prompt response to GRB triggers from satellites in order to measure the early light curves
of GRB optical counterparts (Akerlof et al. 2000b, Akerlof et al. 1999, Kehoe et al. 2001). The normal operation of the ROTSE-I
instrument was completely automatic requiring only periodic maintenance. The telescope consisted of four Canon 200 mm lenses
with f/1.8 focal ratio, each covering $8.2\arcdeg\times8.2\arcdeg$ for a total field of view about
$16\arcdeg\times16\arcdeg$. All four optical elements were carried by a single rapidly slewing mount and
designated with the symbols {\tt a} through {\tt d}. The telescope and mount were located on the roof of a military surplus enclosure,
which housed the instrument control computers, and protected by a clamshell during the day or in bad weather. The instrument was located at the
Los Alamos National Laboratory (W$106\arcdeg 15\arcmin 13\arcsec$, N$35\arcdeg 52\arcmin 9\arcsec$), just outside
the town of Los Alamos, New Mexico. Light pollution at this site, although detectable, had only a minor impact on final survey
photometry. Given the very dry local
climate and 85\% fraction of useful nights it is well suited for survey astronomy. Significant cloud cover is
confined to the monsoon season lasting from July to mid September.

\subsection{Imaging Cameras}
\label{sec:cameras}

Each of the four Canon lenses was equipped with a thermo-electrically (TE) cooled AP-10 camera, which employs a Thomson
TH7899M CCD. The 2k$\times$2k chip format covers an $8.2\arcdeg\times8.2\arcdeg$ degree field of view with 14.4$\arcsec$
pixels. The spatial resolution of the system was limited by instrumental seeing. The Canon lenses delivered a typical point spread
function (PSF) with full width at half maximum (FWHM) of $\sim$20$\arcsec$ and therefore marginally undersampled images of point sources.
The ROTSE-I telescope was operated without any filters so the spectral response is primarily limited by the sensitivity of the CCD,
resulting in a very broad optical band from 450 nm to 1000 nm which covers the photometric bands from mid $B$ to mid $I$.
The quantum efficiency of the front illuminated thick CCD chips makes the effective band most comparable to the Johnson $R$ band.
To optimize the readout speed for GRB response measurements, the images were read in 14-bit mode. There is no loss of information however,
because of the relatively narrow dynamic range of the AP-10 cameras.
The typical gain setting was $\sim$8 e$^{-}$ADU$^{-1}$. Images are sky background limited primarily because of large pixel size.
The limiting $V$ magnitude of the faintest stars recorded in 80-second exposures was typically 14.5--15.5. Saturation occurred
at 10--10.5 mag in normal exposures (80 seconds) and at about 8 mag in bright time exposures (20 seconds). Vignetting in the lenses
is very significant and amounts to about 40\% loss
of sensitivity near the corners of the CCDs. This effect is very stable and easily corrected by flat fielding procedure.
The shutters in the AP-10 cameras did not perform according to their specification. Especially in cold conditions, shutters
did not operate smoothly and caused ``anomalous vignetting'' near the frame edges in some of the images. Photometric
corrections explained in Section~\ref{sec:corrections} remove this and other effects. Dark frames and small scale flat
field features were stable over a few days to a week. A few bad columns in the CCDs did not affect the overall quality of the
data set. Cameras {\tt a}--{\tt d} did not perform equally. The slightly lower photometric quality of camera {\tt d} can be seen
in survey statistics presented in Section~\ref{sec:survey}. The loss of observing time due to temporary failure of camera
{\tt c} is also visible.

\subsection{Observing Protocol}
\label{sec:observing}

Despite the fact that the primary goal of the ROTSE-I project was rapid response to GRB triggers and not sky
patrols, almost all observing time was actually spent in the latter mode. An accessible GRB position would be
posted by the GCN network approximately once every 10 days. Upon receipt of the coordinates, the ROTSE-I system would
abort the current patrol activity and immediately start observing the field around the position for approximately
one hour of imaging.

At the beginning of each night, about 12 dark frames were collected for calibration purposes. No special flat
field exposures were made (Section~\ref{sec:basic}). The large combined field of view delivered by the ROTSE-I system
requires only 206 tiles to cover the entire sky, with 161 tiles observable from Los Alamos. The list of fields
with elevation above 20$\arcdeg$ was prepared by startup scripts. During normal execution of a patrol, for each
of those fields two 80 second exposures would be taken, separated only by the 1.5 minute duty cycle. Reduced exposure
time of 20 seconds was used in bright Moon light conditions ($\sim$30\% of data). No frames were taken when the
moon was closer than 12$\arcdeg$ from the field center. Time keeping accurate to 20 ms was implemented using
Network Time Protocol. On a good night, it was possible to cover the entire local sky ($\sim 10,000$ deg$^2$)
twice. Paired observations are useful for detecting variability and aperiodic transients. They also provide
a handle on spurious detections due to man-made space objects, cosmic rays, hot pixels and other effects.

The position angle of the cameras was fixed at PA=0$\arcdeg$ in all fields except for near-polar region where
the control software allowed PA=$180\arcdeg$. In the latter case, fields normally assigned to cameras {\tt a} and {\tt b}
would be imaged by cameras {\tt c} and {\tt d} respectively. Such observations are flagged appropriately and excluded
from parts of the analysis (Section~\ref{sec:compilation}).

\section{Data Processing}
\label{sec:processing}

Analysis of the data presented in this release was conducted off-line on archival ROTSE-I images. Here we briefly discuss the data reduction
pipeline, schematically shown in Figure~\ref{fig:pipeline}, which was employed to analyze that data.

\subsection{Image Reductions}
\label{sec:image}

\subsubsection{Basic Frame Corrections}
\label{sec:basic}

Between August 1997 and December 2001, ROTSE-I collected 7 TB of image data, however the performance of the system
was not optimal in the first few months of the project and near the end of its lifetime. To build the NSVS, we selected
observations covering 1 full year between April 1999 and March 2000, when the system delivered the best overall data
quality. This limited the raw data set to 225,000 images ($\sim$2 TB). The system would automatically prepare a median
dark frame for each exposure time using all dark images collected on a particular observing night. Dark subtraction
removes a small fraction of pixels ($\ll$1\%) with high dark current rates. Flat field frames were obtained from
a median of all individual patrol images made during a given night. This is possible with a large number of independent
fields ($\sim$80) and statistics limited by sky noise due to large pixel size. Stellar profiles are completely removed
by the procedure. We found that shutter problems (Section~\ref{sec:cameras}) affected some of the flat field images.
Therefore, we visually evaluated all flat field frames and their ratios with frames made on a few other nights
to select calibration sets with consistent large scale properties. As a result, it was possible to correct the vast majority
of the frames using good calibration frames from the same night or the night before.

\vspace{1cm}

\subsubsection{Source Extraction} 
\label{sec:sextractor}

The corrected images are passed to SExtractor software (Bertin \& Arnouts 1996) which reduces them to object lists.
The choice of this source extraction package was motivated by the undersampling of stellar profiles, the significant gradients
of the PSF shape in very wide field images, and overall reduction speed. SExtractor was optimized for reduction of images in galaxy
surveys, but it is known to perform well in moderately crowded stellar fields. In order to optimize sensitivity, our images are
filtered before object detection with a Gaussian kernel employing a FWHM of 2.5 pixels and requiring a minimum of 5 connected
pixels in an object. These basic detections are further thresholded by the software and an attempt is made to break up blended objects.
We use SExtractor aperture magnitudes calculated with the 5 pixel (72$\arcsec$) aperture diameter. Since the ROTSE-I images are
almost completely dominated by stars, we store only a small fraction of information available for each object: position,
magnitude, magnitude error and processing flags. The observed errors at the bright end of the magnitude range are larger
than predicted by simple photon noise. Such discrepancies are common for CCD measurements and are typically generated by residual
systematic effects of flat field errors, thin clouds, PSF variations and sampling. In order to account for these systematic errors,
we had to add, in quadrature, a 0.01 mag contribution to the formal error bars.

\subsubsection{Blending}
\label{sec:blending}

SExtractor does not perform PSF photometry, and in general it is unable to deblend light distributions without a
saddle point. This sets the distance limit of about 3.0 pixels for separation of stellar blends. We found that the default
parameters of the deblender were very conservative, resulting in very large patches of the sky at low Galactic
latitudes being assigned to the same objects. After some experimentation we were able to partially control this process,
however, there are still cases when tight groups of several objects with merging wings are considered to be a single
object. Typically, such aggregates extend up to ten pixels across with a bright object in the middle, but occasionally
in dense parts of the Milky Way, they can be up to 30 pixels wide. This makes the completeness of the NSVS at
$|b|<20\arcdeg$ depend strongly on stellar number density at small spatial scales of around $\sim3\arcmin$. The exact
assignment of pixels to objects may differ from frame to frame and therefore some measurements lost due to severe
blending may still be present among ``orphaned'' measurements, unidentified with any of the light curves
(Section~\ref{sec:compilation}). The term ``orphan'' used here in the context of a photometric detection should not
be confused with orphan GRBs and other optical transients.

\subsubsection{Astrometric and Photometric Matching}
\label{sec:matching}

Initial calibration of object lists consists of transformation of instrumental positions to celestial coordinates and
conversion of raw instrumental magnitudes to a photometric system that can be understood by users. In wide field imaging this
can be done on an image by image basis. Each ROTSE-I field covers 64 deg$^2$ and contains on average about 1500 stars
from the Tycho catalog (Hog 1998). The Tycho catalog, a product of the Hipparcos/Tycho mission, provides very accurate
astrometry and two-color ($B$ and $V$) photometry. Most Tycho stars are fainter than the 10 mag saturation limit of
the ROTSE-I observations. The astrometric matching is performed in the detector plane after
deprojecting the corresponding part of the catalog using a canonical gnomonic projection (Calabretta \& Greisen 2002).
Using an approximate mount position for a given image, the first set of roughly 30 Tycho stars can be identified with
the triangle algorithm. This first order transformation is used to further match a few hundred bright, but
unsaturated, Tycho stars. A third order polynomial warp adequately describes the transformation between the observed
and catalog positions of stars in the detector plane. Finally, transformed (x, y) positions are converted back to
($\alpha_{2000}, \delta_{2000}$) using the inverse of the initial gnomonic projection.

Photometric calibration is somewhat complicated by the very wide unfiltered spectral response of the ROTSE-I imaging system
which spans a large part of the Johnson-Cousins system from mid $B$ to mid $I$ (Section~\ref{sec:cameras}). The best
empirical prediction of a ROTSE magnitude $m_{_{V, {\rm ROTSE}}}$ for Tycho stars is:

$$m_{_{V, {\rm ROTSE}}} = m_{_V} - {{m_{_B} - m_{_V}}\over{1.875}}.$$

\noindent
The median shift between instrumental magnitudes and the above color-corrected magnitudes of Tycho stars is then applied
to all stellar magnitudes from a given image. This procedure puts ROTSE-I measurements onto a $V$-equivalent scale, in the
sense that the mean Tycho star has $m_{_{V, {\rm ROTSE}}} = m_{_V}$. These intermediate object lists are
passed to light curve building software and are subject to further refinements (Section~\ref{sec:corrections}).
Matching to Tycho stars occasionally fails in very crowded areas of the Galactic plane and/or due to substantial cloud
cover. Initial calibration was successful for 184,006 frames.

\subsection{Object Identification and Photometric Corrections}
\label{sec:corrections}

A set of object lists for all exposures of a given field has to be collated in order to identify measurements that
belong to the same objects and hence construct light curves. In the process, a collective look at the temporal behavior of
stars across the field can provide useful information on systematics of the photometry. Very wide field images show some
complications that are usually unimportant in data from narrow field instruments. Pronounced gradients of background,
color dependent atmospheric extinction and gray extinction from thin clouds are common. Color dependent effects cannot
be fully corrected using only a single photometric band. Some ROTSE-I images have additional complications near frame
edges caused by shutter problems (Section~\ref{sec:cameras}). After the processing steps outlined in Section~\ref{sec:image},
most photometric residuals are still correlated over spatial scales of a few hundred pixels. These gradients are handled
by local photometric corrections. The procedure allows measurements of an object originating from each frame
to be expressed on a relative scale with respect to stars in the neighborhood, as given by a master list of stars called
a template. At the same time we can collect diagnostics providing useful measures of the data quality in the final database.
They are used to set the measurement quality flags (Table~\ref{tab:flags}), making it easier to select measurements satisfying
requirements of a particular application. 

\subsubsection{Template Construction}
\label{sec:template}

The process begins with preparation of a template object list for each of the 644 fields. Frames with the {\tt MOUNTFLIP}
flag (see Table~\ref{tab:flags}) or fewer than 1000 detected objects are rejected for this purpose. We also apply a cut on
standard deviation of the position and magnitude offsets around the fit to Tycho stars ({\tt pos\_sigma} $< 0.3$ pixels and
{\tt zp\_sigma} $< 0.4$ mag, see Section~\ref{sec:database}). Measurements of all field objects from all admitted images are
clustered in order to identify persistent objects. To accomplish this, we count the number of individual detections on a simple
grid with pixels equal in size to detector pixels. Local maxima of this histogram containing at least 15\% of all possible
detections are declared as objects. From this counting exercise we derive crude centroids that are further refined using
a 1-pixel identification radius and subsequent sigma clipping. The main properties of objects stored in a template list
for each field are median ($\alpha_{2000}, \delta_{2000}$) positions, median object magnitudes and standard deviation
of individual magnitudes around the median. These are essentially aggregate parameters of the preliminary light curve
for each object. By using median object magnitudes to construct the template, we avoid biasing all field photometry
toward a single frame that may be affected by systematic effects of the kind described above. As a measure
of scatter that is robust against occasional strong outliers, we adopt half size of a centered range that includes 68\%
of all magnitude points.

\subsubsection{Photometric Correction Maps}
\label{sec:maps}

In the next step, template lists are used to derive 2D maps of photometric corrections. The most significant contribution
to systematics in frame-to-frame photometry is caused by clouds, atmospheric extinction gradients and occasional shutter
problems. The photometric errors are therefore correlated over the scales of a few hundred pixels.
By examining a median shift between magnitudes of stars detected in a program frame and template magnitudes for the
same stars in any $200\times200$ pixel area, one obtains a correction to the photometry of stars in the defined
macro-pixel. Stars for which the scatter is more than 4 times the median error bar are rejected as potentially variable.
The frames are then divided into macro-pixels, each covering $200\times200$ detector pixels, to obtain $10\times10$ maps. The map
records relative photometry corrections, the scatter of all magnitude differences in a macro-pixel and the number of objects
available for those calculations. The global scatter of the map is also recorded. All these characteristics can be used
to assess the quality of the data. The pipeline will assign a {\tt null} value to a macro-pixel in cases when there are fewer
than 10 stars to work with, available stars cover less than half of the macro-pixel area, or the value of the correction
or its error are unreasonably high ($>1.0$ mag).

\subsubsection{Compilation of Light Curve Database and Object Catalog}
\label{sec:compilation}

Applying relative photometry corrections and building intermediate database files is the final stage in the data reduction
process. Individual corrections are obtained by bi-linearly interpolating the maps. In the case of a missing macro-pixel
surrounded by valid ones, we allow a patch based on linear interpolation to be applied. Such measurements are flagged
accordingly but were observed to remain reliable. We also flag measurements with large corrections ($>0.1$ mag), large
scatter of magnitude differences used to derive the correction ($>0.2$ mag), and large scatter of all macro-pixels in
the map ($>0.1$ mag). If the patch cannot be computed, there is insufficient information to derive the correction.
This condition sets the {\tt NOCORR} flag signaling a very unreliable measurement. Table~\ref{tab:flags} describes
all the data processing flags. The correction could not be calculated
for fewer than 0.2\% of the database measurements. For most frames 95\% of all correction values are distributed
within $\pm$0.05 mag. In case of a frame affected by a shutter glitch the distribution can be strongly non-Gaussian with the
95\% interval typically extending to $\pm$0.1 mag. In both cases there is a small tail of high correction values extending
to about $\pm$1.0 mag.

Measurements that can be associated with the position of a template object to within 4$\sigma$ are tagged with the object ID
and frame ID and stored in the main light curve database file. The error estimate is available from the template list. There
are no additional restrictions on detections admitted to the main light curve database. The final number of database light
curves is almost 2$\times10^7$ (the same as the number of template objects). There are $\sim$3.35$\times10^9$ individual
observations with identified parent objects. Remaining observations are in the ``orphan'' category. We decided to keep only high
S/N orphan measurements, those with magnitudes 14.5 or brighter and magnitude errors below 0.1. The database contains over
2$\times10^8$ orphan measurements. All measurements are treated individually, in particular joined detection in paired exposures
(Section~\ref{sec:observing}) is not required. Such criteria, based solely on frame epochs, can always be applied to the data
as a post processing step.

At this point we formed a revised catalog of objects with average properties based on the fully processed light curve files.
For each template object we examine a corrected set of measurements and recalculate the median magnitude, magnitude scatter
(significantly improved in most cases), median error bar and other useful characteristics. Only the subset of ``good''
measurements defined in Table~\ref{tab:good} is used in these calculations. These criteria primarily reject error codes
and several flags associated with known problems. Currently, it is a recommended choice for working with NSVS data that offers
a sensible compromise between the amount of data and data quality. Other equally good or better selection cuts may be
possible. A small fraction of all objects below the equator and at the far field edges have fewer than 15 ``good''
measurements and, as such, their data are of limited use. Corresponding catalog entries have light curve statistics copied
from the template list and are flagged as having {\tt TPLSTATS}.

\subsubsection{Multiple Detections in Overlap Regions}
\label{sec:overlaps}

Pointing imperfections in repeated exposures of a given field result in object detections outside the area around the field center
covered by a single CCD format (Table~\ref{tab:specs}). This effectively extends each template by a margin of $\pm150$ pixels
on all sides. Together with similarly sized intended overlap between the fields, this causes multiple detections
of the same physical objects in more than one field. Overlap regions tend to grow closer to the celestial north pole for simple
geometric reasons. About 30\% of the database light curves belong to object references of second order or higher.
In Section~\ref{sec:database} we use those independent light curves to assess systematic errors in positions and magnitudes.
On the other hand, removing multiple references to the same objects requires perfect control over systematic effects. Even if
the influence of shutter problems and atmospheric gradients were known with certainty, some irreducible effects would remain.
The problem is not trivial in the case of unfiltered photometry, especially for fast variable sources, but it becomes much easier
to handle for a few sources of particular interest after the data has been extracted from the database. This is why a global
merge of the physical database has not been performed. We provide a database table with pairs of object IDs that refer to the same
physical object (Section~\ref{sec:database}). Implementation details of this kind can be hidden by the user interface to the database
by defining a mapping between the low level database structures and the top level view where object data appear as merged
(Section~\ref{sec:database}). That approach provides full flexibility to refine the definition of what is the same, and what
is not, as more is known about the data set. 

\section{Public Data Distribution}
\label{sec:data}

\subsection{Survey Coverage and Quality}
\label{sec:survey}

Our Northern Sky Variability Survey covers the entire sky visible from Los Alamos, New Mexico. This includes the entire region north of
the declination $\delta\simeq-38\arcdeg$; more than 30,000 square degrees, 75\% of the celestial sphere. The completeness, quality
of the data and the number of available measurements are noticeably lower at low declinations and low Galactic latitude. The overall
quality of the survey measurements is summarized in Table~\ref{tab:quality}. In Figure~\ref{fig:sky} we show positions of roughly
$5\times10^5$ NSVS stars brighter than 11 mag out of about 14 million stars brighter than 15.5 mag present in the survey. The main
feature in this plot is the overdensity of stars near the plane of the Milky Way. In that region
one can notice relatively less populated areas due to dust lanes. An empty spot near the Galactic Center (field {\tt 156 d}) is an
artifact. The region is basically lost from the survey due to the very low number of available frames. Each dot in Figure~\ref{fig:sky}
represents a time series of typically a few hundred points spread over a total time baseline of 1 year. Temporal coverage is subject
to yearly visibility patterns. The trade-off between this enormous monitoring coverage on one side, and on the other, relatively low
resolution compounded by complexities of very wide field photometry, shaped the final data quality. The faintest objects recorded
in the survey have $V\sim$15.5 mag, however the incompleteness starts increasing sharply at 15th magnitude. Saturation may occur in stars
as faint as 10.5 mag on some nights, but mostly affects stars brighter than 10 mag. Due to exposure times shortened by a factor of 4 during
bright time, some stars as bright as 8.0 mag still have a number of unsaturated measurements sufficient for analysis.

Figure~\ref{fig:stats} summarizes important statistics of the NSVS. Numerous survey parameters strongly correlate with the Galactic latitude
and declination. Several artifacts due to the low volume and quality of available data near the Galactic plane at low declinations are
visible. The bright spot in the plots of number of available frames is due to a special data run in field {\tt 072} (Kehoe et al. 2002).
The darker areas near the Galactic plane are a result of a lower success rate in matching frames to the Tycho catalog. Strong differences near
the celestial pole between panels c) and d) are caused by exclusion of flag {\tt MOUNTFLIP} in the definition of a ``good'' measurement. Lower
than average performance of camera {\tt d} is evident from a periodic pattern in the map of photometric scatter (panel b). The field pattern in
the number of database frames reveals that camera {\tt c} was not collecting data for about 3 months, when it had to be serviced
after an electronics failure. Although the number of useful measurements below the equator drops dramatically to fewer than 100, this
is still sufficient to detect variables and the NSVS remains useful over a large section of the southern hemisphere.

\vspace{1cm}

\subsubsection{Astrometric Errors}
\label{sec:poserr}

Statistical scatter of object positions in individual frames can be better than 0.1 pixel (1.4$\arcsec$) within a single field
(lower part of Figure~\ref{fig:scatter}). Median positions from a large number of measurements should be much more accurate than that,
however that ignores systematic errors of the coordinate system derived separately in each field. A better measure for overall positional
accuracy is the difference of median positions (in 2-D this time) for the same bright unsaturated stars observed in overlapping parts
of adjacent fields. Such differences should be dominated by a systematic contribution and are shown in Figure~\ref{fig:systematics}.
The distribution of these offsets turns out to be comparable to that from Figure~\ref{fig:scatter} and fits well within a single image
pixel. Figure~\ref{fig:systematics} also shows how typical position uncertainties are affected by the high density of stars in the vicinity
of the Galactic plane. The distribution peaks at a higher value for the error and develops a much longer tail due to strong blending.

\subsubsection{Photometric Errors}
\label{sec:photerr}

Figure~\ref{fig:scatter} (upper panel) presents magnitude scatter as a function of median object magnitude in a random field away from
the Galactic equator. Photometric scatter is estimated using ``good'' photometric points (Section~\ref{sec:compilation}) corresponding
to about 75\% of the best data. We consider magnitude scatter for a median star between 11 and 12 mag in each field to be the ``limiting
scatter'', that is the best attainable for a significant fraction of bright stars in a given field. Observations in this magnitude range
are limited by various systematics of photometric conditions and data reductions rather than the statistics of the background. There is
a strong correlation between limiting scatter and Galactic latitude which is evident in Figure~\ref{fig:galb}. The median limiting scatter
over the total survey area is 0.02 mag. However the Galactic plane region, where photometric accuracy suffers significant degradation, has
a higher number density of objects. Averaged over an $8\arcdeg\times8\arcdeg$ field, spatial density of the NSVS objects can vary between 150
and 1000 deg$^{-2}$.

Similar to astrometry, systematics of photometry can be investigated by examining the offsets between magnitudes measured for stars
in overlapping parts of adjacent fields. Histograms of these differences for two random overlap regions at low and high Galactic latitudes
are dominated by constant stars, and are shown in Figure~\ref{fig:systematics}. Half of the stars in the figure differ by 0.04 mag or less
in their median magnitude obtained from light curves constructed independently in neighboring fields. In individual cases, however, such
differences can reach 0.1--0.2 mag. They result primarily from residual shutter problems that propagated to the construction of field
templates, but also from irreducible atmospheric color effects in single broad band photometry. The fact that the histogram width does not
change in a significant way in the proximity of the Galactic equator agrees with our assessment that the differences arise due to systematics
associated with a particular set of field images and not because of high number density of stars or blending. It must be stressed that
the overlap regions between the fields provide the upper bound on the systematic errors. This is where many instrumental effects
(Section~\ref{sec:instruments}) manifest themselves strongest. The internal consistency over the remaining area is certainly much better,
although not easily studied without a suitable external comparison catalog.

\subsection{Public Database}
\label{sec:database}

All photometric time series data in the NSVS is available for public access. The primary means to search and extract the
data is Sky Database for Objects in Time-Domain (SkyDOT\footnotemark[7]; Wo\'zniak et al. 2002). SkyDOT is intended
to become a virtual observatory for variable objects. The site will provide a uniform interface to several large
temporal data sets with a number of time series analysis tools for on-the-spot application to currently viewed data.
SkyDOT is implemented using PostgreSQL\footnotemark[8], an Object Relational Database Management System with support
for practically the entire SQL92 standard. The present database includes six entities listed in Table~\ref{tab:tables}. The columns are explained
in Tables~\ref{tab:columns1} and \ref{tab:columns2}. Five of those tables represent major entities in temporal work: {\tt Field}, {\tt Frame},
{\tt Object}, {\tt Observation}, {\tt Orphan}. The remaining one, {\tt Synonym}, helps to identify multiple references to the same physical
objects (Section~\ref{sec:overlaps}). It implements ``is the same'' relationship between entries of the {\tt Object} table. Users familiar
with SQL can submit queries to the engine with some limitations on the size of output imposed. All users can access the database through
a graphical interface offered by the website. The most popular SQL queries can be accessed through browser buttons and search forms. Currently,
only NSVS photometry is available in the public domain. There is neither a browsing capability nor a direct download option for $\sim$2 TB
of the survey image data. However, image access is planned for future versions of SkyDOT.

\footnotetext[7]{http://skydot.lanl.gov}
\footnotetext[8]{http://www.postgresql.org}

We take advantage of limited precision and range of some database quantities (like magnitudes) and store
them as short integers rather than floating point numbers. This simple rescaling trick generates close to 50\% storage savings
due to a small record size in the main table that dominates the size of the database. Only users of low and intermediate level
data products need to be concerned with these technical details, since they are transparent to top level users who access data
through the presentation layer.

The data is searchable by celestial coordinates. The Hierarchical Triangle Mesh (HTM; Kunszt, Szalay \& Thakar 2001) is used for very rapid
indexing of positions on the sphere. The NSVS database utilizes HTM partitioning of depth 14 with corresponding $\sim$1$\arcmin$ cell size.
Searching for position matches within a small tolerance radius can be performed at a rate of $\sim$20,000 matches per minute. Access time is nearly
uniform across the sphere. In particular, extraction of objects in regions including the north celestial pole is handled seamlessly. HTM runs
as a shared library extension to the database server with SQL wrappers.

\subsection{Other Options for Data Access}
\label{sec:access}

All data discussed in the present release can also be downloaded from the SkyDOT website\footnotemark[7], however transfer of the entire database
over the network may be impractical. Therefore, we will also, upon request, distribute the data on user provided DDS 4 tapes (DAT size). The most
efficient way of storing
the database as files in terms of the size and speed of rebuilding the database proved to be gzip compressed, semicolon separated ASCII
files. To save space, we skip columns with row IDs running from 1 to number of entries. Those can always be deduced. The total size of the
dataset in this form is around 35 GB. It is strongly dominated by the {\tt Observation} table with all object light curves, followed by the
{\tt Orphan} table with observations that could not be identified with any of the objects. Each of the tables {\tt Field}, {\tt Frame} and
{\tt Object} fit within a single file. Tables {\tt Observation} and {\tt Orphan} are broken up into 644 files, one for each field.

\vspace{1cm}

\subsection{Acknowledging NSVS Data}
\label{sec:acknowledging}

We request that the researchers using NSVS in their published work include the following statement to acknowledge the source of data:
\newline

This publication makes use of the data from the Northern Sky Variability Survey created jointly by the Los Alamos National Laboratory
and University of Michigan. The NSVS was funded by the US Department of Energy, the National Aeronautics and Space Administration and
the National Science Foundation.

\section{Discussion and Future Work}
\label{sec:discussion}

We presented the Northern Sky Variability Survey, to this date the most extensive temporal record of the sky on large spatial scales.
All of the survey data is available to the astronomical community and can be searched efficiently using the public SkyDOT database. The database
contains a total of 3.35 billion measurements for approximately 14 million objects in the 8--15.5 magnitude range. Time sampling over one full
year is between twice per night and once every four nights on average. ROTSE-I instrument has achieved a complete spatial coverage of the northern
hemisphere and a large fraction of the southern sky using remarkably low cost hardware. These two factors pose limits to the level of detail
at which variability of the sky was recorded: low spatial resolution, spatial sensitivity variations, a non-standard filter, and complicated systematics
near the Galactic plane.

Despite its limitations, the NSVS is a truly rich source of information on stellar variability. Among stars in the Galaxy, the fraction of variables
with amplitudes detectable by the NSVS is about $\sim$1\% (Eyer 1999, Eyer \& Cuipers 2000). Based on that and on preliminary results in Akerlof et al.
(2000a), one can expect that tens of thousands of new variable stars with good uniform quality light curves are present in the data set. Current database
schema needs to be
expanded along the lines described in Wo\'zniak et al. (2002) to accommodate various types of variables and provide classification capability.
The NSVS combined with astrometric catalogs providing distances and motions, as well as multicolor surveys (2MASS, or even SDSS in a narrow magnitude
range) will enable a comprehensive look at the Galaxy as traced by variable stars. The NSVS objects are bright and therefore the preferred targets for
detailed spectroscopic and astrometric work. Spatial resolution of the survey is not far from that of high energy sky catalogs like the ROSAT
All Sky Survey (RASS; Voges et al. 1999) or the XMM Catalog of Serendipitous Sources (Watson 2003), and therefore well suited for cross correlations.
Perhaps the most exciting questions to be attacked using the NSVS are about rare, hard to find objects. The astronomical
literature provides numerous unexplained reports of variability events on normal stars (e.g. Schaefer, King \& Deliyannis 2000 and references therein).

Photometric monitoring data for Active Galactic Nuclei
(AGN) providing diagnostics of accretion flows is valuable, but limited. Only a few bright AGN are within the magnitude limit of the NSVS
so the real contribution to AGN physics will require deeper flux limits and better resolution in future projects. A major but low cost
improvement in data usability would be the use of a set of standard filters before starting deeper surveys with more frequent time sampling.

Small robotic telescopes with automated data processing pipelines are the best candidates for closing the gap in the current level of temporal
monitoring of the sky. The computing power to perform on-line photometry is available. Experiments like RAPTOR (Vestrand et al. 2002) are
starting to tackle the problem of real time detection and immediate follow-up of short time-scale phenomena. One can envision a monitoring
system capable of partial interpretation of various events occurring on a variety of time-scales and notifying subscribers about interesting changes
of objects in their scientific problem domain. The main challenge is making the immense data stream comprehensible by putting enough smarts into the
software. The sky itself is the ultimate astronomical database that should be mined continuously and in real time. 

\acknowledgments

This work was supported by the Laboratory Directed Research and Development funds at LANL under DOE contract W-7405-ENG-36 to the RAPTOR project
and NASA grant NAG5-5281 to the ROTSE-I collaboration. K.K. was supported by the NSF grant AST-0205813 to Michigan State University,
and S.M. was supported under the auspices of the DOE, NSSA by UC, LLNL under contract W7405-ENG-48.

\clearpage

\begin{deluxetable}{ll}
\tablecaption{\label{tab:specs}{Equipment and Operations in the Northern Sky Variability Survey}}
\tablewidth{14cm}
\tablehead{
\colhead{Parameter} &
\colhead{Value}
}

\startdata
\cutinhead{Telescope and Site}

\makebox[4cm][c]{Geographic position \dotfill} & Los Alamos, New Mexico:
                               W$106\arcdeg 15\arcmin 13\arcsec$, N$35\arcdeg 52\arcmin 9\arcsec$ \\
Elevation           \dotfill & 2300 m \\
Mount speed         \dotfill & 100\arcdeg s$^{-1}$ \\
Telescopes          \dotfill & Four 200mm, f/1.8 Canon lenses \\
Seeing              \dotfill & FWHM$\sim$20\arcsec, instrumental \\
Vignetting          \dotfill & Up to 40\% near frame corners \\
Lens offsets        \dotfill & $\Delta\alpha\cos(\delta)$, $\Delta\delta$ with respect to mount position \\
                             &  \begin{tabular}{rl} {\tt a}: & $-4\arcdeg$,$+4\arcdeg$ \\
                                                    {\tt b}: & $-4\arcdeg$,$-4\arcdeg$ \\
                                                    {\tt c}: & $+4\arcdeg$,$-4\arcdeg$ \\
                                                    {\tt d}: & $+4\arcdeg$,$+4\arcdeg$ \\
                                \end{tabular}  \\

\cutinhead{Imaging Cameras: Four Apogee AP-10s}

CCDs                 \dotfill & 2k$\times$2k Thomson TH7899M chip \\
Gain                 \dotfill & 8 e$^{-}$ ADU$^{-1}$ \\
Read noise           \dotfill & 13--25 e$^{-}$ pixel$^{-1}$ \\
Dynamic range        \dotfill & $>74$ dB \\
Read mode            \dotfill & 14 bit, 1.3 MHz \\
Frame size           \dotfill & 2035$\times$2069 pixel \\
Image scale          \dotfill & 3.50 mm/\arcdeg \\
Pixel size and scale \dotfill & 14$\mu$; 14.4\arcsec pixel$^{-1}$ \\
Filter               \dotfill & Unfiltered optical response $\sim$450nm--1000nm, \\
                              & effective wavelength of $R$ band \\
Field of view        \dotfill & $8.2\arcdeg\times8.2\arcdeg$ per camera \\

\cutinhead{Data collection}

Survey area          \dotfill & 33,326 deg$^2$ ($\delta > -38\arcdeg$),
                                best coverage for $\delta>0\arcdeg$ \\
Time baseline        \dotfill & 1 year, from April 1,  1999 to March 30, 2000 \\
Number of fields     \dotfill & 644=161$\times$4 (cameras {\tt abcd}) \\
Number of frames     \dotfill & 184,006 \\
Number of nights     \dotfill & 275 out of 365 \\
Time sampling        \dotfill & Pairs of frames 1.5 minutes apart, up to 2 pairs per night \\
Calibration          \dotfill & 500--1000 Tycho stars per frame\\

\enddata
\end{deluxetable}

\begin{deluxetable}{crll}
\tablecaption{\label{tab:flags}{Processing Flags}}
\tablecolumns{4}
\tablewidth{18cm}
\tablehead{
\colhead{Hexadecimal} &
\colhead{Decimal} &
\colhead{Flag} &
\colhead{Description}
\\
\colhead{Bit} &
\colhead{Bit} &
&
}
\startdata
\cutinhead{Frame flags}

0x0001 \dotfill & \makebox[1.5cm][c]{\dotfill     1} &  {\tt MOUNTFLIP} & Mount flip near north pole, fields {\tt ab} observed by cameras {\tt cd} \\
0x0002 \dotfill & \dotfill     2 &  {\tt ODDMNTPOS} & Non-standard mount position, only fields {\tt 001 abcd} in year 2000 \\

\cutinhead{Object flags}

0x0001 \dotfill & \dotfill     1 &  {\tt TPLSTATS} & Light curve statistics in {\tt Object} table without photometric corrections\\
0x0002 \dotfill & \dotfill     2 &  {\tt BIGSHIFT} & Final median object centroid more than 1$\sigma$ from preliminary position \\

\cutinhead{Measurement Flags: SExtractor}

0x0001 \dotfill & \dotfill     1 &  {\tt NEIGHBORS} & $\geq$10\% of object area affected by a neighboring object or bad pixels \\
0x0002 \dotfill & \dotfill     2 &  {\tt BLENDED}   & Object is a result of deblending procedure \\
0x0004 \dotfill & \dotfill     4 &  {\tt SATURATED} & Object has at least 1 saturated pixel \\
0x0008 \dotfill & \dotfill     8 &  {\tt ATEDGE}    & Object is truncated by image boundary \\
0x0010 \dotfill & \dotfill    16 &  {\tt APINCOMPL} & Aperture data incomplete or corrupted \\
0x0020 \dotfill & \dotfill    32 &  {\tt ISINCOMPL} & Isophotal data incomplete or corrupted \\
0x0040 \dotfill & \dotfill    64 &  {\tt DBMEMOVR}  & Memory overflow occurred during deblending \\
0x0080 \dotfill & \dotfill   128 &  {\tt EXMEMOVR}  & Memory overflow occurred during extraction \\
\cutinhead{Measurement Flags: Photometric Correction}

0x0100 \dotfill & \dotfill   256 &  {\tt NOCORR}    & Relative photometry correction could not be calculated \\
0x0200 \dotfill & \dotfill   512 &  {\tt PATCH}     & Map of relative photometry corrections was patched to derive correction \\
0x0400 \dotfill & \dotfill  1024 &  {\tt LONPTS}    & Low number of points in a macro-pixel ($<$10) \\
0x0800 \dotfill & \dotfill  2048 &  {\tt HISCAT}    & High scatter of magnitude differences in macro-pixel ($>$0.2 mag) \\
0x1000 \dotfill & \dotfill  4096 &  {\tt HICORR}    & High value of correction ($>$0.1 mag) \\
0x2000 \dotfill & \dotfill  8192 &  {\tt HISIGCORR} & High scatter of corrections across the map ($>$0.1 mag) \\
0x4000 \dotfill & \dotfill 16384 &  {\tt RADECFLIP} & Mount flip near north pole occurred (fields {\tt 001}--{\tt 032 abcd} only) \\
\enddata
\end{deluxetable}

\begin{deluxetable}{ll}
\tablewidth{16.5cm}
\tablecaption{\label{tab:good}{Definition of a Good Photometric Point}}
\tablehead{
\colhead{Condition} &
\colhead{Serves primarily to remove}
}
\startdata
$5.0<$ magnitude $<16.0$            \dotfill  & SExtractor error codes \\
$0.0<$ mag error $<0.4$             \dotfill  & SExtractor error codes \\
\makebox[4.5cm][c]{{\tt !SATURATED} \dotfill} & Saturated measurements; important for stars in and out of saturation \\
                   {\tt !NOCORR}    \dotfill  & Generally unreliable macro-pixels \\
                   {\tt !LONPTS}    \dotfill  & Measurements from far frame edges or shallow images \\
                   {\tt !HISCAT}    \dotfill  & Macro-pixels with unreliable photometry \\
                   {\tt !HICORR}    \dotfill  & Macro-pixels substantially affected by cloud cover or shutter problem \\
                   {\tt !HISIGCORR} \dotfill  & Frames substantially affected by cloud cover or shutter problem \\
                   {\tt !RADECFLIP} \dotfill  & Possible camera dependent systematics; harmless in many applications \\
\enddata
\end{deluxetable}

\begin{deluxetable}{ll}
\tablewidth{18.5cm}
\tablecaption{\label{tab:quality}{NSVS Data Quality}}
\tablehead{
\colhead{Parameter} &
\colhead{Value}
}
\startdata
\makebox[4cm][c]{Pointing offsets \dotfill} & $\sigma\sim0.3\arcdeg$ (75 pixels) per coordinate \\
PSF                    \dotfill & FWHM$\sim$1.5 pixels, undersampled, spatially variable, temporally stable \\
Saturation             \dotfill & 10--10.5 mag, up to 8 mag in bright time \\
Limiting magnitude     \dotfill & $\sim$15.5 mag, 14.5 mag in bright time \\
Astrometric errors     & \\
\hspace{1cm}     Random\tablenotemark{a} \dotfill & magnitude dependent, $1\sigma$=0.7--4.3$\arcsec$,
                                                    $1\sigma$=1.4--5.8$\arcsec$ at Galactic $|b|<20\arcdeg$ \\
\hspace{1cm} Systematic\tablenotemark{b} \dotfill & median deviation $\sim$1.2$\arcsec$ in general field,
                                                    $\sim$2.5$\arcsec$ at Galactic $|b|<20\arcdeg$ \\
Photometric errors     & \\
\hspace{1cm}     Random\tablenotemark{a} \dotfill & limiting scatter $1\sigma$=0.02 mag in median field,
                                                    $1\sigma$=0.02--0.05 mag at Galactic $|b|<20\arcdeg$ \\
\hspace{1cm} Systematic\tablenotemark{b} \dotfill & median deviation $\sim$0.04 mag near frame edges, up to 0.2 mag in extreme cases, \\
                                                  & improving near field center \\
Blending               \dotfill & Stars closer than 3 pixels (84$\arcsec$) generally merged, \\
                                & severe at Galactic $|b|<20\arcdeg$, $\sim$5\% loss of survey area \\
Time sampling          \dotfill & Twice per night to once every $\sim$4 nights \\
Number of epochs       \dotfill & $\sim$200 for average light curve, follows yearly visibility across the sky \\
Number of objects      \dotfill & $\sim$14 million \\
\enddata
\tablenotetext{a}{frame to frame within the same field}
\tablenotetext{b}{systematic difference for the same object in overlap region between fields}
\end{deluxetable}

\begin{deluxetable}{lrl}
\tablecaption{\label{tab:tables}{Database tables}}
\tablewidth{17.5cm}
\tablehead{
\colhead{Table} &
\colhead{Number of rows} &
\colhead{Description}
}
\startdata
\makebox[2.5cm][c]{{\tt Field} \dotfill} & \dotfill           644 & ROTSE-I patrol tiles, each camera counted separately \\
{\tt Frame}       \dotfill & \dotfill       184,006 & Image header and post-processing frame quality information \\
{\tt Object}      \dotfill & \dotfill    19,995,106 & One record of aggregate information for each light curve in the database \\
                           &                        & counting separately same object detections from different fields \\
{\tt Synonym}     \dotfill & \dotfill    14,582,566 & Pairs of light curve IDs referring to the same physical object \\ 
{\tt Observation} \dotfill & \dotfill 3,353,171,900 & All measurements for all light curves \\
{\tt Orphan}      \dotfill & \dotfill   208,106,474 & Measurements unidentified with any of the light curves \\
                             &                        & but brighter than 14.5 mag and with errors $<0.1$ mag \\
\enddata
\end{deluxetable}

\begin{deluxetable}{lrll}
\tablecaption{\label{tab:columns1}{Explanation of table columns: tables Field and Frame}}
\tablewidth{17cm}
\tablehead{
\colhead{Column Name} &
\colhead{Data Type} &
\colhead{Unit} &
\colhead{Description}
}
\startdata
\cutinhead{{\tt Field} Table}

{\tt  id}        \dotfill & {\tt int32}    & \makebox[1.5cm][c]{\dotfill} & Field ID (primary key) \\
{\tt  name}      \dotfill & {\tt char[4]}  & \dotfill & Field name \\
{\tt  rac}       \dotfill & {\tt float32}  & \dotfill deg      & $\alpha_{2000}$ of field center \\
{\tt  decc}      \dotfill & {\tt float32}  & \dotfill deg      & $\delta_{2000}$ of field center \\
{\tt  glc}       \dotfill & {\tt float32}  & \dotfill deg      & Galactic $l$ of field center \\
{\tt  gbc}       \dotfill & {\tt float32}  & \dotfill deg      & Galactic $b$ of field center \\
{\tt  nobs}      \dotfill & {\tt int32}    & \dotfill & Number of frames \\
{\tt  nobj}      \dotfill & {\tt int32}    & \dotfill & Number of catalog objects \\
{\tt  sig\_ph}   \dotfill & {\tt float32}  & \dotfill mag      & Limiting photometric scatter \\

\cutinhead{{\tt Frame} Table}

{\tt  id}         \dotfill & {\tt int32}    & \dotfill & Frame ID (primary key) \\
{\tt  field\_id}  \dotfill & {\tt int32}    & \dotfill & Field ID (foreign key) \\
{\tt  fname}      \dotfill & {\tt char[20]} & \dotfill & Image file name \\
{\tt  camera}     \dotfill & {\tt char[2]}  & \dotfill & Camera ID ({\tt abcd}) \\
{\tt  mjd}        \dotfill & {\tt float64}  & \dotfill day      & $JD-2450000.5$ \\
{\tt  obstime}    \dotfill & {\tt float64}  & \dotfill s        & Time of observation in UT seconds \\
{\tt  date\_obs}  \dotfill & {\tt char[22]} & \dotfill & UT date of observation \\
{\tt  exptime}    \dotfill & {\tt float32}  & \dotfill s        & Exposure time \\
{\tt  bkg}        \dotfill & {\tt float32}  & \dotfill counts   & Sky background \\
{\tt  bkg\_sigma} \dotfill & {\tt float32}  & \dotfill counts   & Standard deviation of sky background \\
{\tt  pos\_sigma} \dotfill & {\tt float32}  & \dotfill pixels   & Standard deviation around the fit to positions of Tycho stars \\
{\tt  zp\_offset} \dotfill & {\tt float32}  & \dotfill mag      & Median magnitude offset with respect to Tycho stars \\
{\tt  zp\_sigma}  \dotfill & {\tt float32}  & \dotfill mag      & Standard deviation of magnitude offsets \\
{\tt  m\_lim}     \dotfill & {\tt float32}  & \dotfill mag      & Limiting magnitude \\
{\tt  sat\_mag}   \dotfill & {\tt float32}  & \dotfill mag      & Saturation magnitude \\
{\tt  nobj\_det}  \dotfill & {\tt int32}    & \dotfill & Number of objects detected by SExtractor \\
{\tt  nobj\_ext } \dotfill & {\tt int32}    & \dotfill & Number of objects actually measured \\
{\tt  nmatch}     \dotfill & {\tt int32}    & \dotfill & Number of Tycho stars used in magnitude matching \\
{\tt  dmoon}      \dotfill & {\tt float32}  & \dotfill deg      & Angular distance between frame center and the Moon \\
{\tt  elev}       \dotfill & {\tt float32}  & \dotfill deg      & Elevation of frame center with respect to horizon \\
{\tt  azimuth}    \dotfill & {\tt float32}  & \dotfill deg      & Azimuth of frame center \\
{\tt  mount\_ra}  \dotfill & {\tt float32}  & \dotfill deg      & $\alpha_{2000}$ position of telescope mount \\
{\tt  mount\_dec} \dotfill & {\tt float32}  & \dotfill deg      & $\delta_{2000}$ position of telescope mount \\
{\tt  offst\_ra}  \dotfill & {\tt float32}  & \dotfill deg      & $\Delta\alpha\cos(\delta)$ offset between frame center and mount position \\
{\tt  offst\_dec} \dotfill & {\tt float32}  & \dotfill deg      & $\Delta\delta$ offset between frame center and mount position \\
{\tt  map\_rms}   \dotfill & {\tt float32}  & \dotfill mag      & Standard deviation of the photometric correction map \\
{\tt  map\_npix}  \dotfill & {\tt int16}    & \dotfill & Number of valid pixels in photometric correction map \\
{\tt  flags}      \dotfill & {\tt int16}    & \makebox[1.5cm][c]{\dotfill} & Frame flags \\
\enddata
\end{deluxetable}

\begin{deluxetable}{lrll}
\tablecaption{\label{tab:columns2}{Explanation of table columns: tables Object, Synonym, Observation and Orphan}}
\tablewidth{17cm}
\tablehead{
\colhead{Column Name} &
\colhead{Data Type} &
\colhead{Unit} &
\colhead{Description}
}
\startdata
\cutinhead{{\tt Object} Table}

{\tt  id}         \dotfill & {\tt int32}   & \makebox[2.5cm][c]{\dotfill} & Object ID (primary key) \\
{\tt  rao}        \dotfill & {\tt float64} & \dotfill deg      & Median $\alpha_{2000}$ of object centroid \\
{\tt  sig\_rao}   \dotfill & {\tt float64} & \dotfill deg      & Standard deviation of individual $\alpha_{2000}$ positions \\  
{\tt  deco}       \dotfill & {\tt float64} & \dotfill deg      & Median $\delta_{2000}$ of object centroid \\
{\tt  sig\_deco}  \dotfill & {\tt float64} & \dotfill deg      & Standard deviation of individual $\delta_{2000}$ positions \\  
{\tt  htm\_id}    \dotfill & {\tt int64}   & \dotfill & HTM ID for quick spatial queries on the sphere \\
{\tt  mag}        \dotfill & {\tt float32} & \dotfill mag      & Median object magnitude from ``good'' points \\
{\tt  rms\_mag}   \dotfill & {\tt float32} & \dotfill mag      & Standard deviation of ``good'' points around median \\
{\tt  med\_err}   \dotfill & {\tt float32} & \dotfill mag      & Median error bar of ``good'' points \\
{\tt  n\_obs}     \dotfill & {\tt int16}   & \dotfill & Number of ``good'' points \\
{\tt  n\_noflip}  \dotfill & {\tt int16}   & \dotfill & Number of all points without {\tt RADECFLIP} flag \\
{\tt  n\_points}  \dotfill & {\tt int16}   & \dotfill & Number of all object detections \\
{\tt  flags}      \dotfill & {\tt int16}   & \dotfill & Object flags \\

\cutinhead{{\tt Synonym} Table}

{\tt  id1}        \dotfill & {\tt int32}   & \dotfill     & First object ID (composite primary key) \\
{\tt  id2}        \dotfill & {\tt int32}   & \dotfill     & Second object ID (composite primary key) \\
{\tt  separation} \dotfill & {\tt float32} & \dotfill deg & Spherical distance ($<$1 pixel) \\

\cutinhead{{\tt Observation} Table}

{\tt  obj\_id}    \dotfill & {\tt int32} & \dotfill    & Object ID (composite primary key \& foreign key) \\
{\tt  frame\_id}  \dotfill & {\tt int32} & \dotfill    & Frame ID (composite primary key \& foreign key) \\
{\tt  dra}        \dotfill & {\tt int16} & \dotfill 1/32000 deg & $\alpha_{2000}$ offset from median centroid \\
{\tt  ddec}       \dotfill & {\tt int16} & \dotfill 1/32000 deg & $\delta_{2000}$ offset from median centroid \\
{\tt  mag}        \dotfill & {\tt int16} & \dotfill mmag        & Corrected object magnitude \\
{\tt  err}        \dotfill & {\tt int16} & \dotfill mmag        & Magnitude error \\
{\tt  flags}      \dotfill & {\tt int16} & \dotfill    & Measurement flags \\

\cutinhead{{\tt Orphan} Table}

{\tt  frame\_id}  \dotfill & {\tt int32}   & \dotfill & Frame ID (composite primary key \& foreign key) \\
{\tt  rao}        \dotfill & {\tt float64} & \dotfill deg      & $\alpha_{2000}$ of object (composite primary key) \\
{\tt  deco}       \dotfill & {\tt float64} & \dotfill deg      & $\delta_{2000}$ of object (composite primary key) \\
{\tt  mag}        \dotfill & {\tt int16}   & \dotfill mmag     & Corrected object magnitude \\
{\tt  err}        \dotfill & {\tt int16}   & \dotfill mmag     & Magnitude error \\
{\tt  flags}      \dotfill & {\tt int16}   & \dotfill & Measurement flags \\
{\tt  htm\_id}    \dotfill & {\tt int64}   & \makebox[2.5cm][c]{\dotfill} & HTM ID for quick spatial queries on the sphere \\

\enddata
\end{deluxetable}

\clearpage

\includegraphics{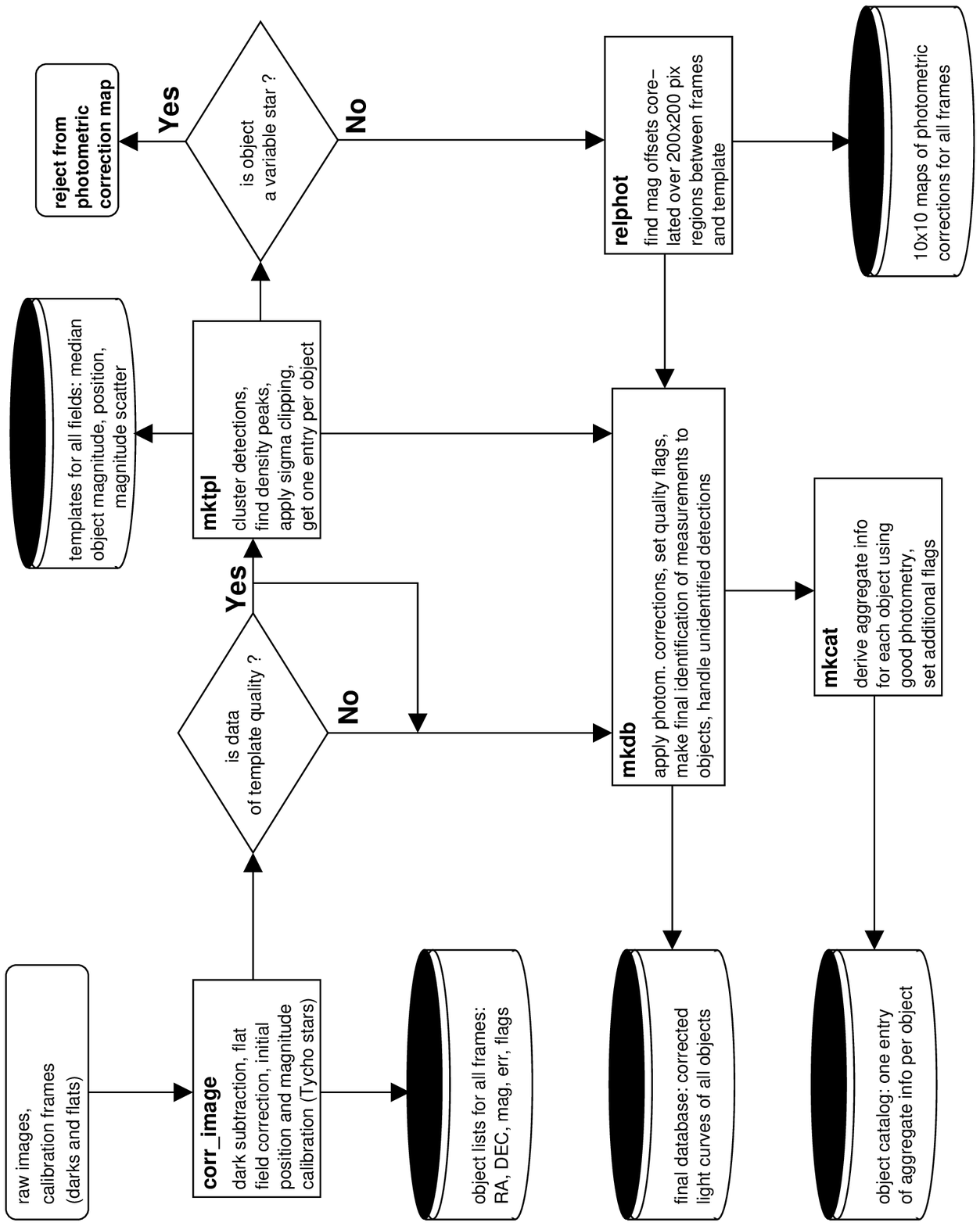}

~~
\vspace{14cm}
\figcaption[figure1.ps]{The NSVS data processing pipeline. \label{fig:pipeline}}

\vspace{3cm}
\begin{center}
{\LARGE This figure is available in gif format: figure2.gif}
\end{center}
\vspace{3cm}

\figcaption[figure2.ps]{Positions of NSVS objects brighter than 11 mag (about 500,000) in equal area Mollweide's
projection. Each dot represents an NSVS object with a temporal history typically composed of a few hundred measurements
covering the 1 year baseline. \label{fig:sky}}

\clearpage

~~
\vspace{4cm}
\begin{center}
{\LARGE This figure is available in gif format: figure3.gif}
\end{center}
\vspace{4cm}

\figcaption[figure3.ps]{NSVS at a glance. Four panels show all sky gray scale maps smoothed over 8$\arcdeg$ spatial scale:
a) Number density of NSVS objects, b) photometric scatter for bright unsaturated stars calculated using ``good''
measurements, c) median number of points per light curve, d) median number of ``good'' photometric points per
light curve. \label{fig:stats}}

\vspace{4cm}
\begin{center}
{\LARGE This figure is available in gif format: figure4.gif}
\end{center}
\vspace{4cm}

\figcaption[figure4.ps]{Random errors as given by frame to frame scatter in a typical field.
Photometric errors (upper) and position errors (lower) are shown as a function of median object magnitude. \label{fig:scatter}}

\clearpage

\includegraphics{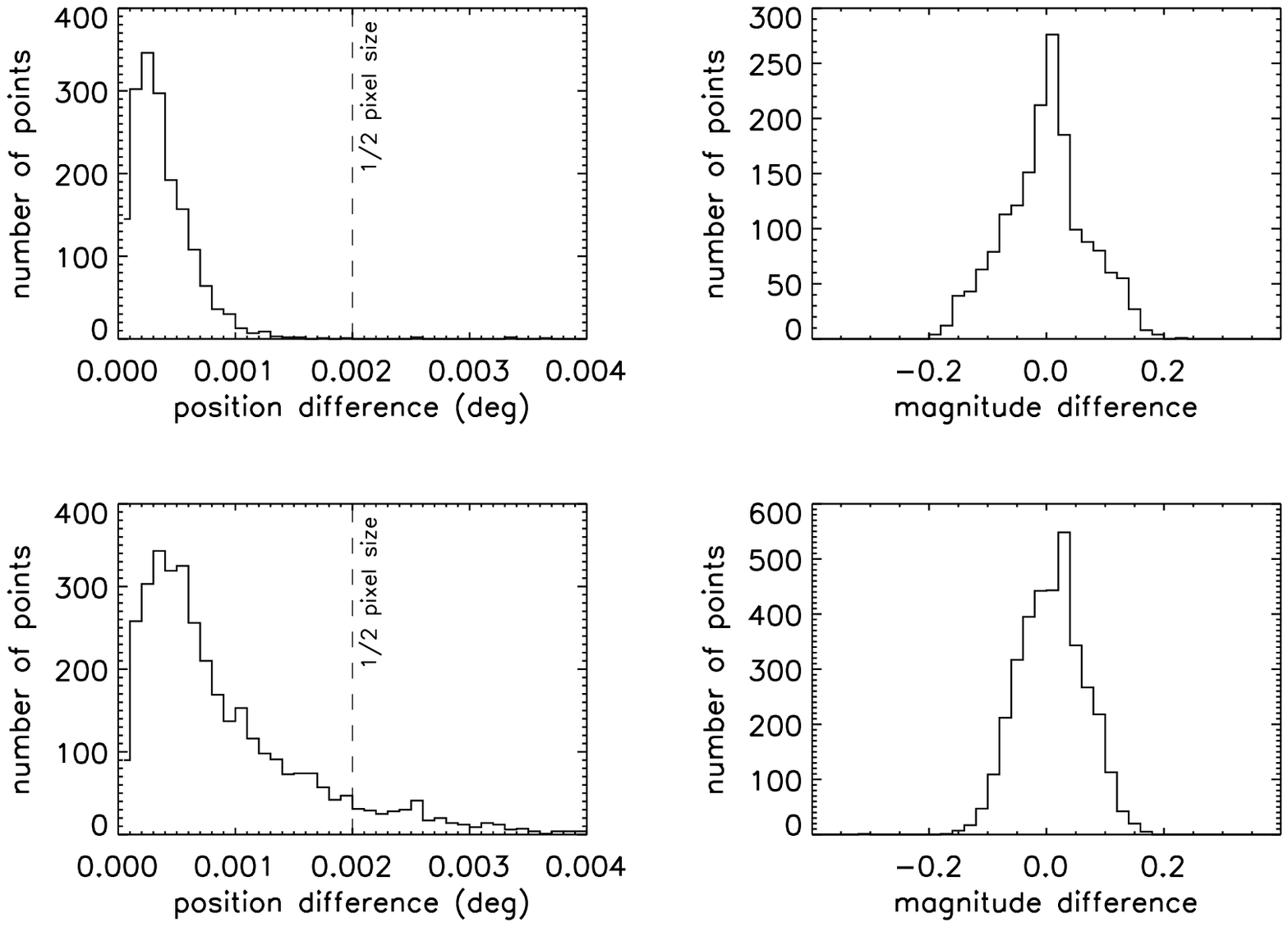}

~~
\vspace{12cm}
\figcaption[figure5.ps]{Systematic errors as given by differences between multiple detections
of the same objects in overlap regions between the adjacent fields in the Galactic plane (lower) and near
the Galactic pole (upper). Shown are differences in median object positions (left) and median object magnitudes
(right) for bright unsaturated stars. Position differences fit well within a small fraction of a pixel. Magnitude
offsets result from residual shutter problems and properties of very broad band photometry in the presence of
intrinsic spread of object colors. Estimates based on overlap regions, where numerous instrumental effects are strongest,
provide an upper bound on systematic errors. \label{fig:systematics}}

\includegraphics{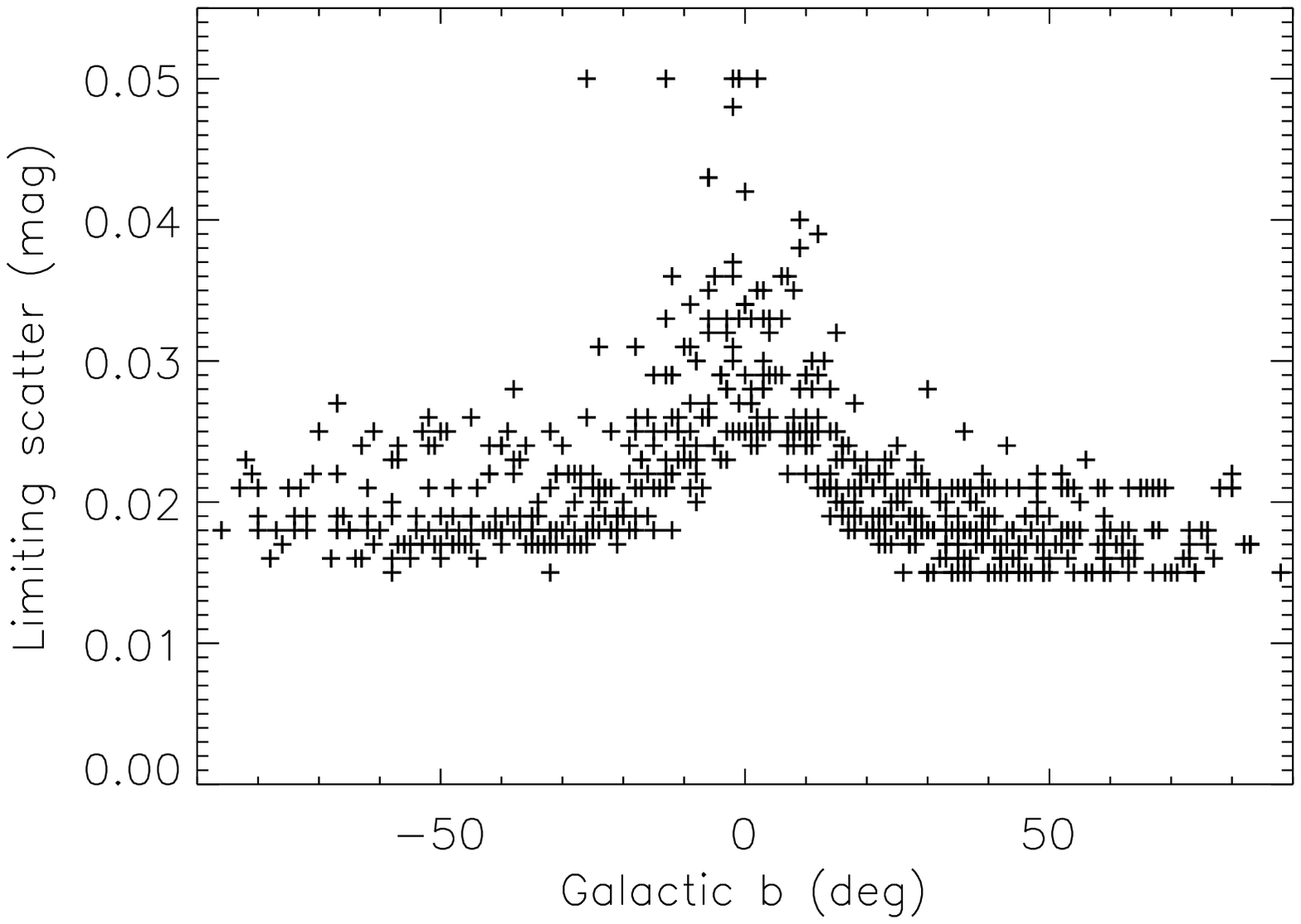}

\vspace{8.5cm}
\figcaption[figure6.ps]{Limiting photometric scatter as a function of the Galactic latitude. In each field
we show light curve scatter for bright unsaturated stars calculated using ``good'' measurements. \label{fig:galb}}

\end{document}